# A statistical analysis of drug seizures and opioid overdose deaths in Ohio from 2014 to 2018

Lam Tran[1], Lin Ma[1], and David White[1]

[1]Denison University, Granville, OH, USA

**Abstract:**

*Objective.* To examine the association between drug seizures and drug overdose deaths in Ohio from 2014 to 2018.

*Methods.* We use linear regression, ARIMA models, and categorical data analysis to quantify the effect of drug seizure composition and weight on drug overdose deaths, to quantify the lag between drug seizures and overdose deaths, and to compare the weight distributions of drug seizures conducted by different types of law enforcement (national, local, and drug task forces).

*Results.* Drug seizure composition and weight have strong predictive value for drug overdose deaths ($F = 27.14$, $p < 1e-15$, $R^2 = .7799$). A time series analysis demonstrates no statistically significant lag between drug seizures and overdose deaths or weight. Histograms and Kolmogorov-Smirnov tests demonstrate stark differences between seizure weight distributions of different types of law enforcement ($p < 1e-7$ for each pairwise comparison).

*Conclusions.* Consideration of drug composition and weight can inform law enforcement seizure activity. To save lives, law enforcement should emphasize seizures of low weight drugs that contain fentanyl.

## 1. Introduction

The number of drug overdose deaths quadrupled in the US from 16,849 cases in 1999 to 70,273 cases in 2017 (NIH, 2019). According to the National Center for Health Statistics (NCHS), in 2018, more than 130 people died every day in the US after overdosing on opioids (CDC/NCHS, 2018). Synthetic opioids, especially illegally made fentanyl and its analogues, is hypothesized to be the main driver of overdose deaths, as they were involved in 67.8% of the cases and the percentage is still rising (CDC, 2018). In 2017, Ohio had the second-highest rate of drug overdose deaths (32.4 deaths per 100,000), and similarly to the national opioid epidemic, synthetic opioids were the leading cause (3,523 deaths, over 82%; NIH/NIDA, 2019). Unfortunately, toxicology data from drug overdose deaths often lags by several months, so that authorities only see an influx of dangerous fentanyl analogues once it is too late to prevent the deaths these analogues cause. A new dataset, of law enforcement searches and seizures of drugs, has a much lower lag, and offers the opportunity for an "early warning system" to alert authorities to the presence of dangerous fentanyl analogues in the drug supply, before these analogues cause an increase in drug overdose deaths (Rosenblum et. al, 2020).

      Our research uses two data sources: one is the dataset of drug seizures from Ohio's Bureau of Criminal Investigation's crime lab (BCI), while the other is the unintentional drug overdose death dataset from the Ohio Public Health Information Warehouse. Both datasets run from 2014 to 2018. The BCI dataset contains information on confiscation date and county, the weight of the drug seizure, which law enforcement organization made the seizure, and the chemical composition of the seizure. Section 2 provides a detailed illustration of both datasets. Sections 3-6 build a sequence of statistical models that layer in more and more aspects of the BCI data. Section 3 begins with the simplest model: we use the Ohio Department of Health Mortality dataset (CDC, 2018) to chart the relationship between fentanyl seizures and deaths over time, to explore the degree to which seizures lag behind deaths, and to quantify the amount of variability in deaths that it explained by drug seizures and by fentanyl seizures. This model ignores information about the weights of drug seizures, about the time-series nature of the data, and about the type of law enforcement unit that made the seizure. However, this simplistic analysis (easily communicated to non-statistician policymakers) is enough to confirm that fentanyl is driving the opioid epidemic, and to quantify the extent to which law enforcement drug

seizures involving fentanyl (and its analogs) predict drug overdose deaths. That is, we quantify the efficacy of drug seizure data as an "early warning system" (Rosenblum et. al, 2020) and are able to make concrete policy recommendations, e.g., that all crime labs should share their data, and that the resulting information about the prevalence of fentanyl in the drug supply be shared with law enforcement, harm reduction professionals, and drug users. We note that, in many cases, deaths involving fentanyl also involved other drugs, suggesting that fentanyl is being mixed into the drug supply in ways that can surprise drug users (e.g., mixtures of fentanyl with methamphetamines, or fentanyl with cocaine, as well as pure opioid overdoses such as fentanyl mixed with heroin) and hence lead to an increased number of deaths.

In Section 4, we introduce the weight variable from the BCI dataset. We show that low weight drug seizures are more likely to contain fentanyl than higher weight seizures. We then fit a linear model of different weight bins to predict overdose deaths. This model shows that small weight seizures explain more variation in death than higher weight seizures. In Section 5, we fit a time-series ARIMA model (Shumway and Stoffer, 2005) to our data, to improve on the model in Section 3 by factoring the past into our model. This ARIMA model includes the weight variable, building on the model of Section 4. In Section 6, we compare the efficacy of different types of law enforcement, including national law enforcement (e.g., FBI/DEA), drug task forces, and local police and sheriff's departments, as regards the drug weights they routinely seize. Our goal in this section is not to improve on the model from Section 5, but rather to shine light on a different aspect of law enforcement activity. We present both histograms and Kolmogorov-Smirnov tests to show that different types of law enforcement seize very different weight distributions. Local police departments seem to be best at seizing low weight drugs, which, as we present in Section 4, are more dangerous. We conclude the paper with a summary of our main results, including policy recommendations for law enforcement and harm reduction professionals. Specifically, while seizures of all sizes remove drugs from the market, an important consequence of our analysis is that, to save lives, focus at the end of the drug pipeline is needed, rather than only at the source. It is essential that drug users be informed about the presence of potent fentanyl analogues adulterating the drug supply, and that efforts by law enforcement and harm reduction professionals aim to prevent drug users from ingesting such adulterated drugs.

## 2. Background and related works

In the recent years, countless scholars have researched the opioid epidemic from a multitude of perspectives. In a review article published in 2017, Armenian et al. explained the timeline of the main synthetic opioid events of the past 50 years, then discussed the clinical effects of fentanyl (and its analogs), such as central nervous system depression and overdose death. Fentanyl is a synthetic opioid pain reliever that is approved for treating severe pain, such as cancer. It is an extremely concentrated opioid that is 50 times more potent than heroin, and can be cheaply and easily synthesized (Beletsky, et. al, 2017). Recent data collected by the Center for Disease Control and Prevention indicated that the number of fentanyl encounters more than doubled, from 5,343 in 2014 to 13,882 in 2015. According to the iron law of prohibition, when law enforcement is strict, drug traffickers prefer potent, high value drugs such as fentanyl, which can be delivered in smaller shipments (Beletsky, et. al, 2017). As fentanyl is difficult to dose correctly, these small batches can quickly result in mass deaths (Beletsky, et. al, 2017). In recent years, deaths due to prescription pills (semi-synthetic opioids) and heroin have leveled off, but deaths due to fentanyl have increased sharply (Ciccarone, 2017). For this reason, we focus in this paper on drug seizures and deaths where fentanyl was involved. Our companion paper focuses on traits of individuals who died due to unintentional drug overdose, identifying trends in the interests of maximizing the impact of harm reduction work (Ma, Tran, White 2024).

According to Armenian, little is known about the potency of several fentanyl analogs (Armenian, 2017). Armenian et al. also comprehensively reviewed fentanyl, fentanyl analogs, and novel synthetic opioids, pointing out that potent synthetic opioids evolve quickly as soon as they are regulated, making detection and prevention from these drugs harder over time (2017). Consequently, throughout this paper, when we speak of deaths due to fentanyl (or fentanyl seizures), we mean fentanyl and any of its analogues.

Our use of time-series analysis, and our focus on the efficacy of law enforcement, builds on several related papers. One such paper approaches the epidemic by demonstrating 50 state alcohol and drug agencies' robust initiatives being used to deal with the crisis (Wickramatilake et al., 2017). Another used time series analysis to evaluate the impact of overdose education and naloxone distribution programs on opioid overdose rate in Massachusetts (Wally et al., 2013). Finally, Livingstons et al. also used time series analysis to focus on the change in opioid overdose death after recreational cannabis was legalized in Colorado (2017). Despite the scope of the opioid epidemic in Ohio, there is a dearth of published papers addressing the crisis in Ohio. There have been a number of government reports (from the Ohio Department of Health, the National Institute on Drug Abuse, the Recovery Ohio Advisory Council, and from the CDC), of unpublished academic studies (e.g. by the Swank Program in Rural-Urban Policy), and blog posts by Harm Reduction Ohio (https://www.harmreductionohio.org/) that show the situation of the crisis here, but very few published, academic papers addressing Ohio specifically. This paper remedies the gap in the literature, analyzing the opioid epidemic in Ohio from several new angles.

The first dataset we analyzed is from Ohio Bureau of Criminal Investigations (BCI) crime lab, from 2014 to 2018. We obtained this dataset from Harm Reduction Ohio, a non-profit organization, and our project was vetted by Denison University's Institutional Review Board. BCI had no role in our research, and this paper in no way represents the views of BCI. The BCI dataset contains information about each drug seizure by law enforcement in the period of January 2014 to December 2018, with records from testing physical drug samples and residues taken from drug paraphernalia. The information includes, for each drug seizure, the unique identifier of the seizure, the types of drugs confiscated, the weight of the drugs, the county where the seizure occurred, the date of the seizure, and the name of the law enforcement department that performed the seizure. Since our data on deaths is only available at the level of month and year, we converted the date into month and year. This dataset is not a random sample of drug seizures in Ohio. Rather, it consists of every single drug seizure BCI analyzed in 2014-2018. There are seven other crime labs in Ohio that analyze drug seizures, and we do not have access to their data. Despite this limitation, our analysis demonstrates that the BCI dataset has strong predictive power for drug overdose deaths, and we hope that publicizing this power will lead to the other crime labs making their data publicly available.

All of our analyses were carried out using the statistical computing language R. Because Harm Reduction Ohio provided us with two separate datasets, one with the observations from 2014 to 2017 and the other with data from 2017 to 2018, we first cleaned and merged these datasets. The cleaning process involved renaming columns to be consistent, filtering out missing data cases, identifying duplicate cases, and filtering out observations with unclear results (e.g.

unknown substances, or cases where testing was not done). These steps resulted in a well-formatted data frame with 12 columns and 139,076 observations, which we have included as supplemental data along with this paper. Of these observations, 118,936 of them (85%) have data on the weight of the confiscated drugs tested by BCI. The columns consists of information such as Identify number, Date, Lab Case Number, Lab Item Number, County, Department Name, Drugs Type, Weight, Month, Year, Combine number and Duplicate. The weight variable refers to the weight that BCI tested, and this is not always the same as the weight of the drug seizure (i.e. it is possible for BCI to test only a sample of the seizure). However, even with this limitation, we demonstrate that weight has strong power for predicting overdose deaths, and we encourage BCI to begin to report the weight seized along with the weight tested. Among the 139,076 observations, 14,382 seizures involved fentanyl or its analogs (12%). We created a separate data frame that only includes these 14,382 seizures, and we refer to this data frame as the "fentanyl seizures." In Section 3, we demonstrate that fentanyl seizures have better predictive value for overdose deaths than all seizures. This makes sense, because the data frame of all seizures contains entries like cannabis seizures, which are not as lethal as fentanyl.

The second dataset we used is a publicly available dataset from the Ohio Public Health Information Warehouse that was accessed on May 22, 2019 (This is the same dataset as in the CDC WONDER database). With the help of the Website's built-in report creator, we selected the report "Number of Resident Deaths by Year, Ohio", filtered for only unintentional drug poisoning from 2014-2018 with year and month as variables. In other words, this dataset provides us information about the monthly number of overdose deaths of all different types of drugs in Ohio from 2014-2018. We looked at all overdose deaths, rather than only those caused by fentanyl or only those caused by opioids, for several reasons. First, from a harm reduction standpoint, the most important response variable is all overdose deaths. Secondly, there was concern about duplicate entries in overdose deaths due to opioids, because an individual who died with multiple opioids in his or her bloodstream would be present in different subsets from the overdose death dataset. Thirdly, there was concern that an opioid user might have died from using non-opioid drugs, especially if police seizures of opioids had increased. Lastly, the model we selected, of using fentanyl seizures to predict all overdose deaths, had stronger predictive value than using more general drug seizures or predicting overdose deaths due to just fentanyl or just opioids (these results are reported in Section 3). In this way, our model provides us a better understanding of the overall circumstances of the drug fatalities instead of just a portion of it.

In order to use seizure data to predict overdose deaths, we merged the two datasets. Since overdose deaths are only reported per month, all of our models required us to aggregate seizure data to the level of months, e.g. considering the total number of fentanyl seizures in a month, or the number of seizures of a certain weight in a month. We avoided working with average weight or total weight seized, because our analysis in Section 4 shows that lower weight seizures are more dangerous than higher weight seizures (because lower weight seizures have a larger percentage containing fentanyl), and summing or averaging the weight variable destroys this important difference.

## 3. Relationship between Seizures and Death

Our initial analysis was inspired by a report from Harm Reduction Ohio called *Carfentanil's deadly roles in Ohio overdose deaths* (Cauchon, 2018). Extending this analysis to include data from 2018, we obtain the following graph (Figure 1), which uses time plots to chart the trends of fentanyl seizures, carfentanil seizures, and overdose death in Ohio each month from 2014 to 2018.

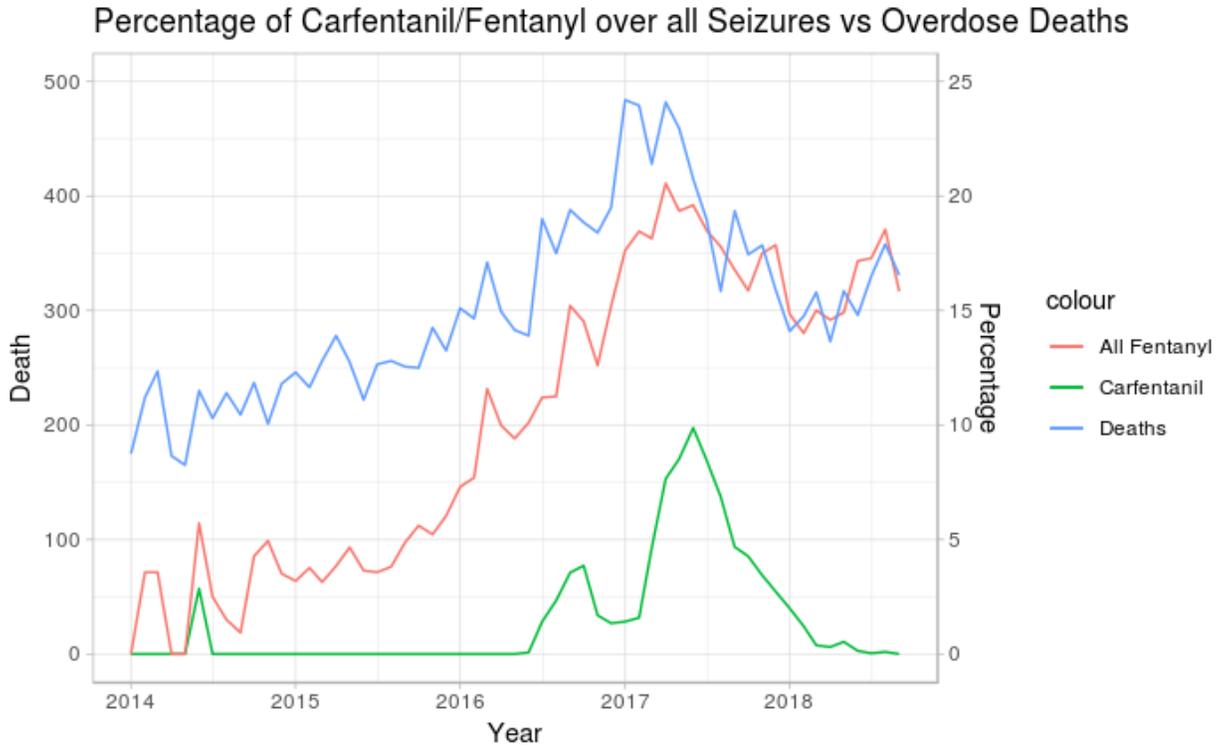

Figure 1: Line plot of percentage of carfentanil and fentanyl seizures with overdose death.

Cauchon uses this graph (in the range 2014-2017), and another analysis (Cauchon, 2017), to study the lag between when deaths spike and when the presence of carfentanil spikes. He quantifies the lag as 3 months, obtained as the number of months between the maximum point on the death time series and the maximum point on the carfentanil time series. In this section, we prewhiten the data, then use the statistical tool of the cross-correlation function (standard in time-series analysis; Shumway and Stoffer, 2005 and Bahid et. al. 2024) to quantify the lag between deaths and seizures over the entire time series, rather than only at the peaks. This function returns the correlation between one time series (deaths), and lags of another time series (seizures). Both positive and negative lags are analyzed. We found that fentanyl seizures (which, recall from Section 2, include analogs of fentanyl such as carfentanil) has a stronger correlation with overdose deaths than carfentanil seizures do. Furthermore, we found that there is no statistically significant lag (except lag 0) between overdose deaths and fentanyl seizures. In other words, fentanyl seizures do not lag behind deaths (nor do deaths lag behind fentanyl seizures), lending credibility to the hypothesis that seizures provide an early warning system for deaths, since seizure data is available on a faster time scale.

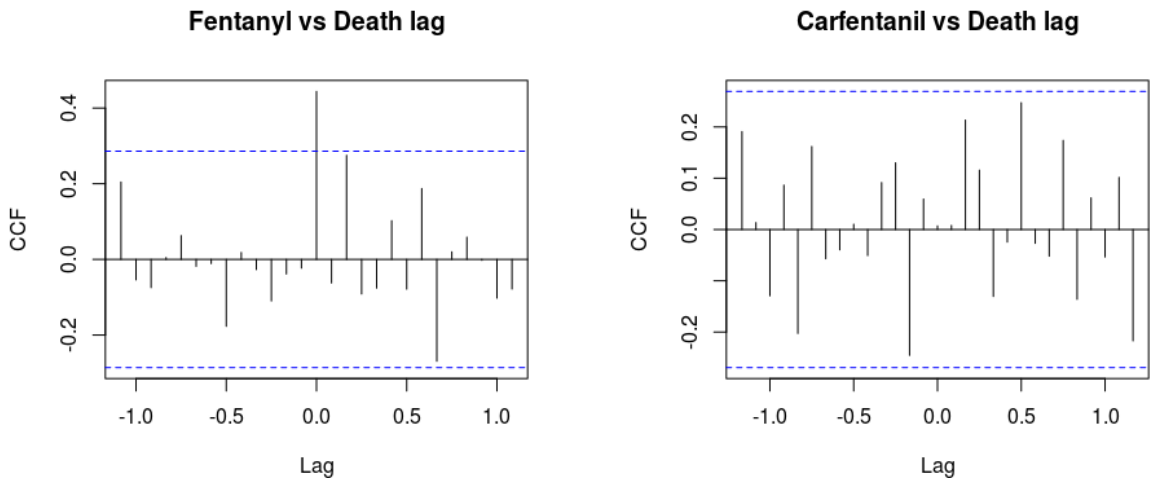

Figure 2: Cross correlation plot for lag between overdose deaths and Carfentanil/Fentanyl Seizures.

Figures 1 and 2 focus on fentanyl seizures and carfentanil seizures separately because Cauchon wanted to draw attention to the danger of carfentanil in particular (Cauchon, 2018). However, now that we have identified the outlier role of carfentanil in 2017, we will focus on fentanyl in the aggregate (i.e., fentanyl and all its analogs), which includes carfentanil. We first construct a linear regression model for overdose deaths with respect to the number of fentanyl seizures, and the $R^2$ value of this model gives us one way to measure the extent to which fentanyl seizures predict death. In order to fit a linear model, we needed several transformations. First, we needed to take the log of fentanyl seizures, to respond to skewness in the data caused by a few months in which many fentanyl seizures occurred, and a few months where very few fentanyl seizures occurred. As the scatterplots below show, the relationship is not linear, even after taking the log, so our model includes a quadratic term. Further transformations resulted in patterns in the residuals and/or lower adjusted $R^2$. Using all drug seizures, instead of fentanyl seizures, resulted in a lower $R^2$ value (of 0.3386) even after transformations were applied to achieve a linear relationship. This is likely because many drug seizures consist of drugs that are not deadly, and this justifies our focus on fentanyl seizures. Using opioid drug seizures (except for the opioids used in medically assisted treatments for opioid use disorder) also resulted in a lower $R^2$ value (of 0.2855), again because many opioid drug seizures that do not contain fentanyl are not as dangerous. For the remainder of the paper, whenever we discuss BCI seizures, we are focusing only on seizures that contain fentanyl or its analogs.

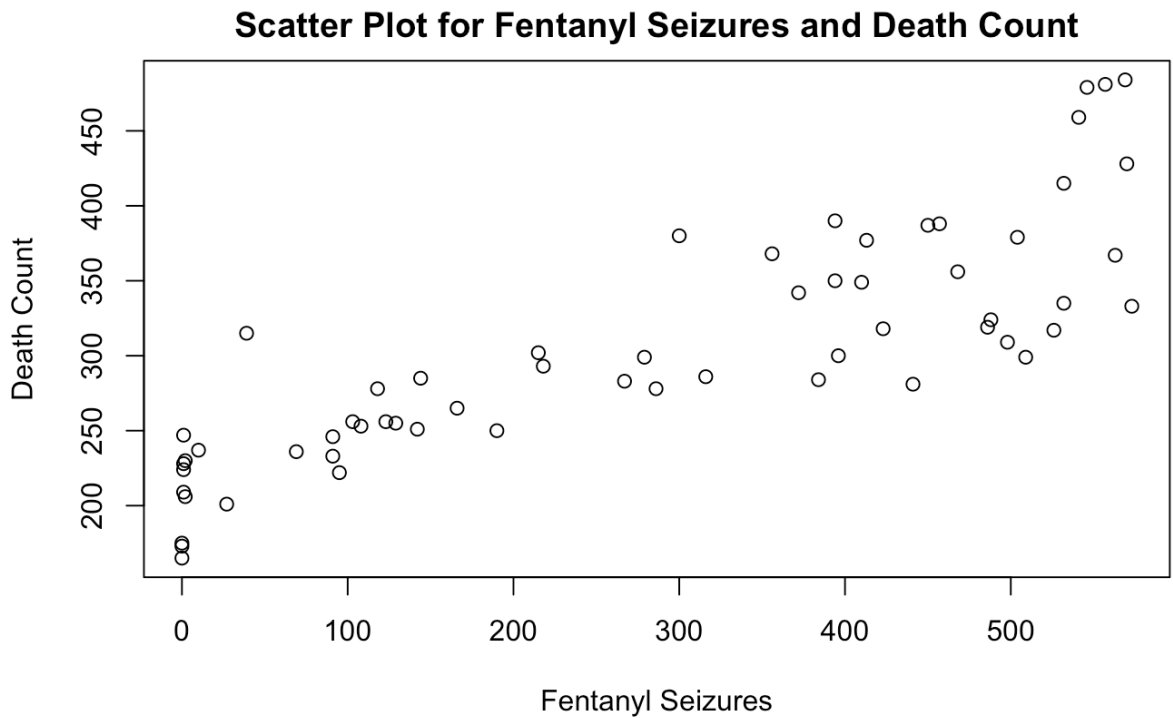

**Figure 3: Scatter Plot for Fentanyl Seizures and Death Count**

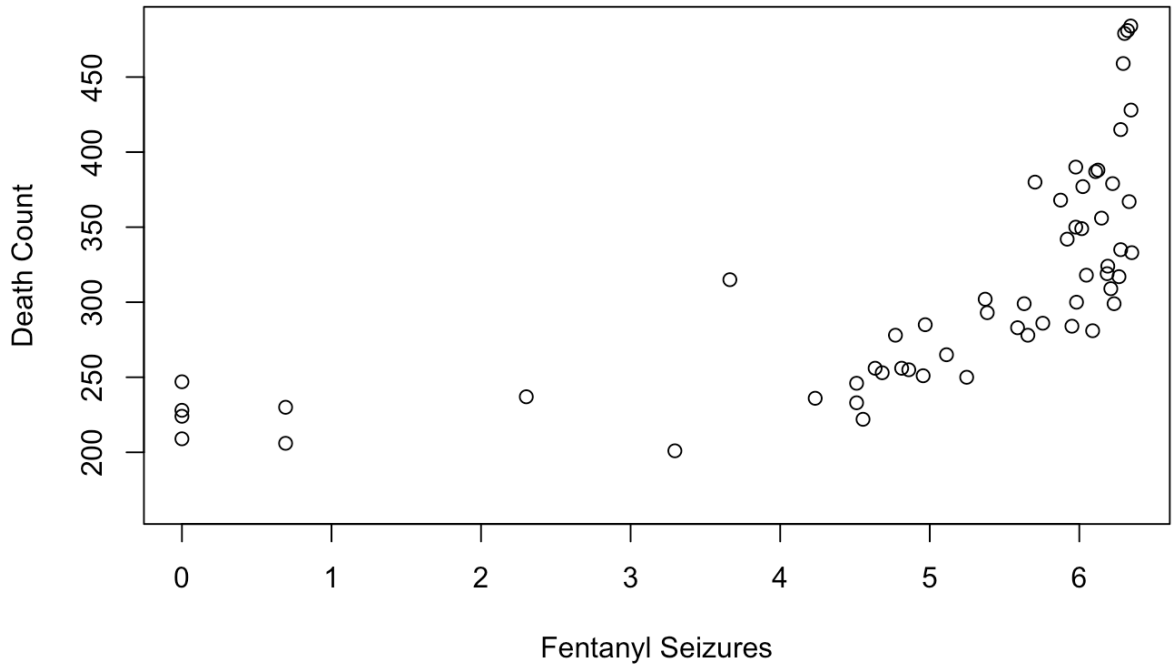

Figure 4: Scatter Plot for logged Fentanyl Seizures and Death Count.

To summarize, we fit the following model for deaths and fentanyl seizures:

$$\text{Deaths}_t = \beta_0 + \beta_1 * \log(\text{Fent}_t) + \beta_2 * \log(\text{Fent}_t)^2 + \varepsilon$$

where $\text{Deaths}_t$ (resp. $\text{Fent}_t$) refers to the number of deaths (resp. fentanyl seizures) in month $t$, and $\varepsilon$ is the error term. This model is statistically significant (F = 45.11; p < 1e-11). Both Figure 4 and the values of $\beta_1$ and $\beta_2$ demonstrate that, as the number of fentanyl seizures increases, so does the number of drug overdose deaths. This may seem counter-intuitive, as more seizures reduce the supply of drugs available to drug users. However, having more fentanyl seizures also implies that there is more fentanyl in the drug supply, and as Figure 1 conclusively demonstrates, when there is more fentanyl in the drug supply, more drug users will die. The residuals of this model demonstrate that there is no further pattern, and are homoscedastic. Since the data consists of 60 months, normality for the residuals ($\varepsilon$) is not required in order to interpret the significance of the coefficients. The adjusted R² value of this model is **0.599**, which tells us that about 59% of the variance in overdose deaths is explained by fentanyl seizures.

This model relates drug seizures and overdose deaths each month, but does not factor in the past. Because each data point represents a month, there is the possibility of autocorrelation. We address this in Section 5 by treating **Deaths** and **Fentanyl** as time-series, and fitting an ARIMA model. The resulting model improves on the model in this section. Nevertheless, even without the tools from time-series analysis, the model above provides strong evidence that law enforcement seizures of fentanyl are strongly predictive for deaths (in the same month). Before introducing the tools of time-series analysis, we address the importance of the weight variable.

## 4. The importance of weight

In this section, we examine whether low weight seizures are more dangerous or deadly than high weight seizures. In order to use weight to predict monthly deaths, we must aggregate the weight variable to the monthly level. For this, we followed an analysis by Cauchon and divided the weight into 11 bins, with the thresholds of 0.1, 0.24, 0.5, 1, 2, 5, 10, 20, 50, and 100 grams (Cauchon, 2019). We then counted the number of seizures in each bin for each month. We used 11 dummy variables to tell which bin a given seizure fell into. Cauchon's analysis suggested that cocaine seizures of low weight were more likely to contain fentanyl than higher weight seizures (Cauchon, 2019). Figure 5 and Table 1 confirm this relationship for all drugs (rather than only for cocaine). Note that most seizures containing fentanyl have weight less than 100 grams. This suggests fentanyl is being mixed into the drug supply later in the drug pipeline, and that law enforcement cannot ignore low weight seizures if the goal is to save lives.

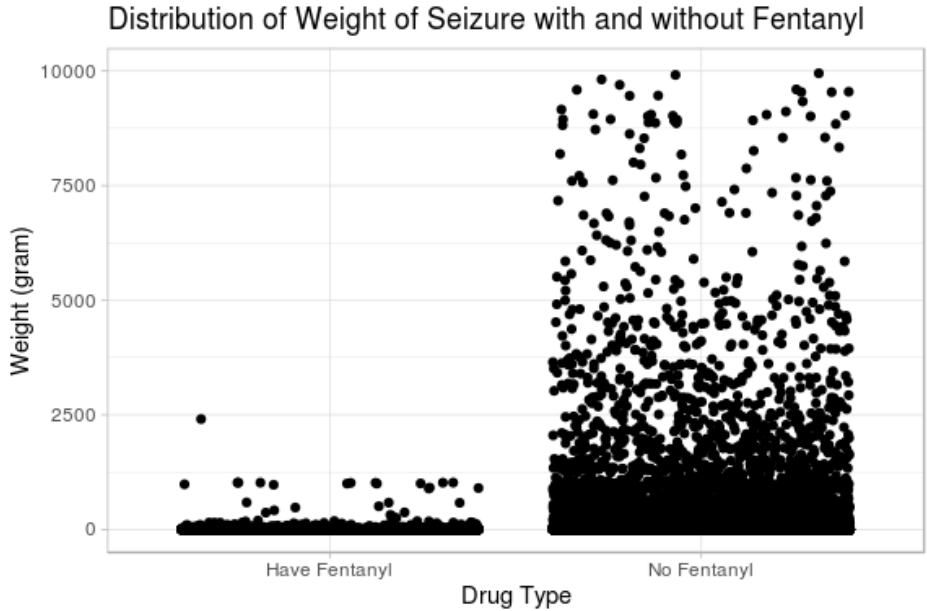

**Figure 5: Comparison between distribution of weight of seizures with and without fentanyl.**

Based on Figure 5 and Table 1 (below), one might predict that the number of low weight seizures is a better predictor for deaths than the number of high weight seizures, as low weight seizures are more likely to contain fentanyl. Indeed, we found that each of our 11 dummy variables, in isolation, was a statistically significant predictor for death ($p < 0.001$), and that more seizures predicted for more deaths. The lowest category (0-0.1g) has the highest explanatory power for death ($R^2 = 0.4686$). The category (0.1-0.24g) has $R^2 = 0.3858$, while the category (0.24-0.5g) has $R^2 = 0.4035$. In aggregate, the first six bins (representing drug seizures of weight 0 to 5 grams) yield an adjusted R-squared of **0.6427**, while adding higher weight bins decreases the value of adjusted R-squared. When adding the number of fentanyl-related seizures in each month to the model, and including the seventh bin (0 to 10 gram), the adjusted R-squared value jumps to **0.7799**. Including higher weight bins decreases the adjusted R-squared value. Even in the presence of the weight variables, the coefficient for fentanyl remained positive, i.e. a larger number of fentanyl seizures predicts for more deaths each month.

| Size of seizure, net weight | All seizures | Fentanyl Found | % seizures with fentanyl | $R^2$ |
|---|---|---|---|---|
| Above 100 grams | 4512 | 57 | 1.26% | 0.2213 |
| 50-100 grams | 1682 | 63 | 3.74% | 0.2164 |
| 20-50 grams | 3715 | 164 | 4.41% | 0.2210 |
| 10-20 grams | 3706 | 242 | 6.53% | 0.2810 |
| 5-10 grams | 4883 | 362 | 7.41% | 0.2541 |
| 2-5 grams | 8778 | 763 | 8.69% | 0.2606 |
| 1-2 grams | 8154 | 852 | 10.45% | 0.2670 |
| 0.5 -1 gram | 13080 | 1531 | 11.70% | 0.3436 |
| 0.24-0.5 gram | 22430 | 2337 | 10.42% | 0.4136 |
| 0.1-0.24 gram | 24115 | 2869 | 11.90% | 0.3963 |
| < 0.1 gram | 23892 | 5143 | 21.53% | 0.4776 |

Table 1: Fentanyl adulteration by weight bins and $R^2$ of each bin as a separate predictor for deaths

We conclude this section with a word on the interpretation of the significance of the coefficients in our best model (using the first seven bins, and fentanyl seizures, to predict deaths; adjusted $R^2$ = 0.7799). Because the 11 dummy variables together make up all seizures, including more than one indicator variable of weight in any model will induce multicollinearity. Consequently, the magnitude and sign of the individual coefficients, and their statistical significance, should not be interpreted in such models (Neter, et. al. 1996). The results in this section support our hypothesis that low weight seizures are the most deadly, and reinforce the results from Section 3 that fentanyl seizures have a significant impact on overdose deaths. As in Section 3, all of the regression assumptions except are autocorrelation are satisfied by this model. The concrete policy recommendation is that law enforcement should focus on low weight seizures, and that information about the composition of low weight seizures each month should be shared with harm reduction professionals (and, eventually, with drug users) so that users will not be surprised by fentanyl adulterating the drugs they ingest. When data on low weight seizures suggests a larger than normal prevalence of fentanyl or a dangerous analogue such as carfentanil, harm reduction tools--such as narcan, fentanyl test strips, and direct communication with drug users at needle exchanges--should be deployed to save lives.

## 5. Dependence on the past

In the previous sections, we concentrated on analyzing the immediate impact of drug seizures on overdose deaths. In this section, we introduce an ARIMA (Autoregressive Integrated Moving Average) model that factors in the effect of the past as well, and corrects autocorrelation issues. Similar analyses have been carried out for protest data (Rodríguez and White, 2023) yielding conclusions regarding protests in the USA (White, Rodríguez, and Topaz, 2024), in Ukraine (Bahid, et. al., 2024), and in Georgia (White, et. al., 2024). Fitting an ARIMA model consists of finding three numbers $p,q,r$ relating the present in a time series to its own past, and then checking that the assumptions of the model are satisfied (Shumway and Stoffer, 2005). The first number, $p$, is the degree of *autoregressive* dependence on the past. For us, this means the number of previous months that the current month depends on, for our time series of drug overdose deaths. The second number, $q$, is the number of times the time series must be *differenced* in order to be *stationary*. The precise meanings of these terms is not necessary for the reader to understand at the moment, but it is important that the final model be stationary in order to correctly interpret the model coefficients and their *p*-values. The third number, *r*, is the *moving-average* dependence on the past. Basically, this is the number of previous months that the current number of deaths depends on, but now it is dependence on the error terms in the model, rather than on the actual number of deaths in the previous months. It is standard in time-series analysis to use the *autocorrelation function* (ACF) and *partial autocorrelation function* (PACF) to determine the numbers *p* and *r*, and to use the augmented Dickey–Fuller test to determine the number *q* . It is also standard to use the AIC and BIC values to compare different choices of models (Shumway and Stoffer, 2005). Figure 6 below shows the ACF and PACF for our time series of drug overdose deaths, and they are consistent with *p* = *1* and *r* = *1*. This choice of *p* and *r* also yielded the best AIC and BIC values.

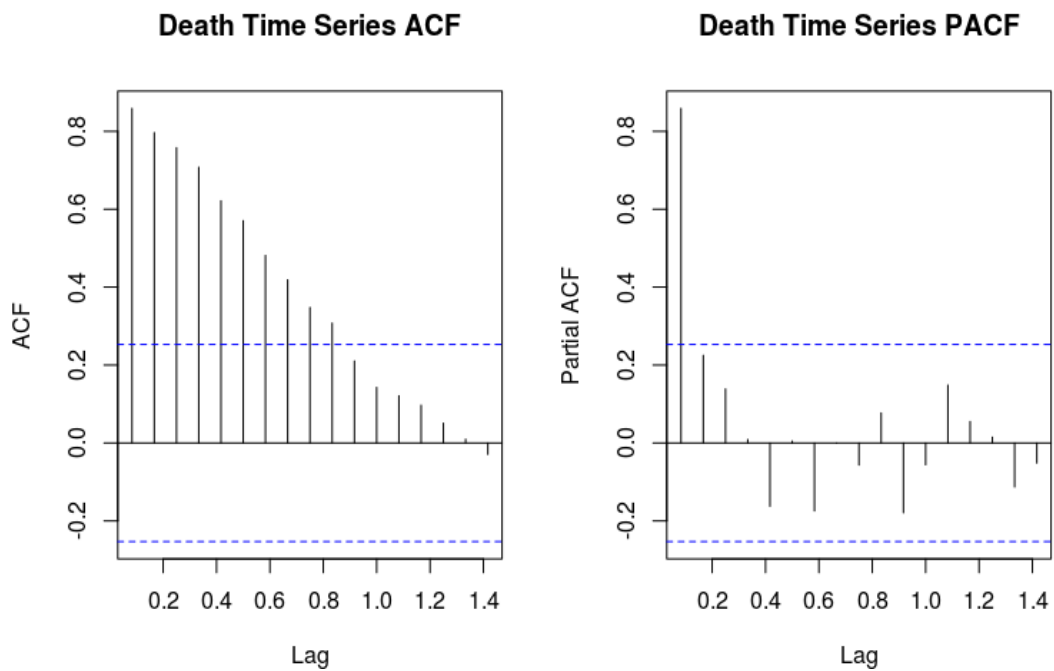

**Figure 6: The plots of the autocorrelation and partial autocorrelation functions of the death time series.**

The augmented Dickey–Fuller test (ADF) test is a hypothesis test in which the null hypothesis states the time series is not level or trend stationary and the alternative hypothesis states that the time series is level or trend stationary. When we ran the test with $q = 2$, we obtained a p-value of 0.01 which (together with the ACF and PACF above) suggested an ARIMA(1,2,1) model, and we included this model in the published version of this paper. However, subsequent analysis suggested that the ARIMA(1,2,1) model was over-differenced, and an ARIMA(1,1,0) model is better. Residual analysis demonstrates that first-order differencing is sufficient. Note that the selection of a model with degree 1 differencing represents an assumption that the time series has a trend that varies over time, and Figure 1.1 demonstrates that this assumption is justified. The model additionally assumes the residues are uncorrelated and normally distributed. We first address normality, via a QQ plot.

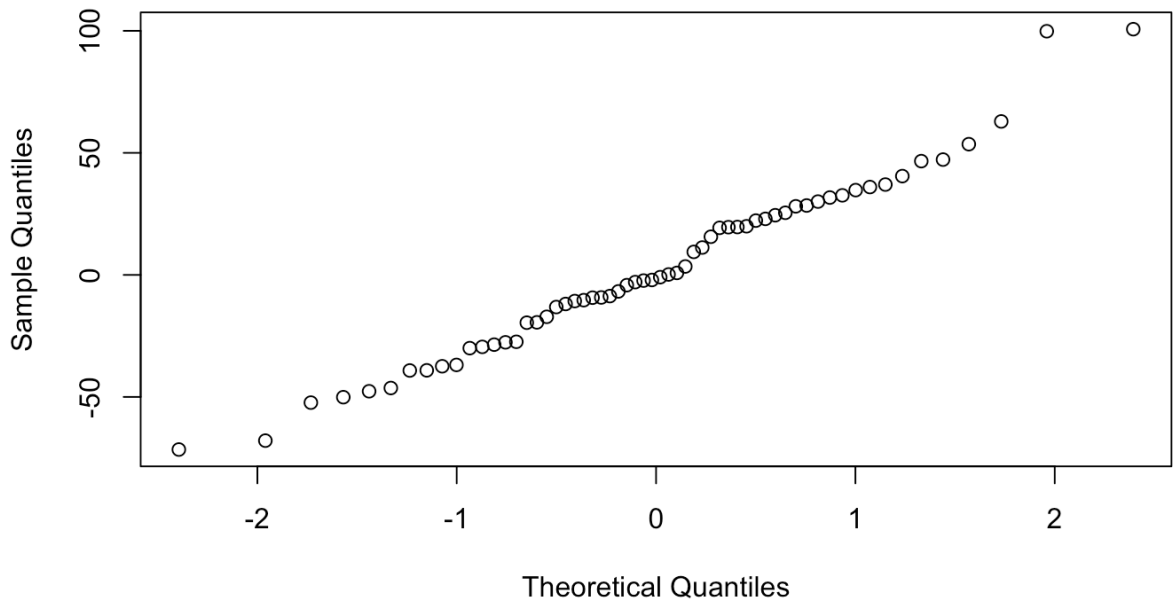

Figure 7: QQ plot of the residuals of the model ARIMA(1,1,0) on death time series.

Most of the points lie on a straight line with few outliers, suggesting that residuals are approximately normally distributed. Since there are 60 months, approximate normality of the residuals is sufficient to deduce normality of the coefficients and to interpret p-values. Lastly, inspection of the residual plot suggests it contains no further trend, i.e., residuals are uncorrelated, as required in an ARIMA model.

We build on this ARIMA model, and on the models of the previous sections, by fitting a model for overdose deaths as a function of the number of fentanyl seizures and the number of seizures in certain weight bins. Based on the log likelihood and AIC, the best ARIMA model is:

**Model 3.1: Deaths$_t$ = ARIMA(1,1,0) + $\beta_1$*Fent$_t$ + $\beta_2$*Weight$_t^{0to0.1}$+ $\beta_3$*Weight$_t^{0.1to0.24}$ + ε**

where **Fent$_t$** is the number of fentanyl seizures in month t, and **Weight$_t^{0to0.1}$** (resp. **Weight$_t^{0.1to0.24}$**) is the number of seizures in weight bin 0-0.1 grams (resp. 0.1-0.24 grams) in month t. Recall that this model expands on the models in Sections 3 and 4. All coefficients are statistically significant (p < 0.05) except for $\beta_3$, and as we discussed in Section 4, this should not be surprising due to multicollinearity between the weight variables. For instance, when we drop the variable **Weight$_t^{0to0.1}$**, we find that **Weight$_t^{0.1to0.24}$** does have a significant impact on death (p < 0.001). In Section 3, we fit a model for death based on **Fent** alone, and an inspection of residuals revealed that the relationship was not linear. However, Model 3.1 has more variables, which affects the relationship between **Fent** and **Death**. Both AIC calculations and residuals show that it is best to include **Fent** rather than **log(Fent)** or a quadratic term. Hence, Model 3.1 improves on the models of Sections 3 and 4, and is also easier to interpret, assuming familiarity with the techniques of time-series analysis.

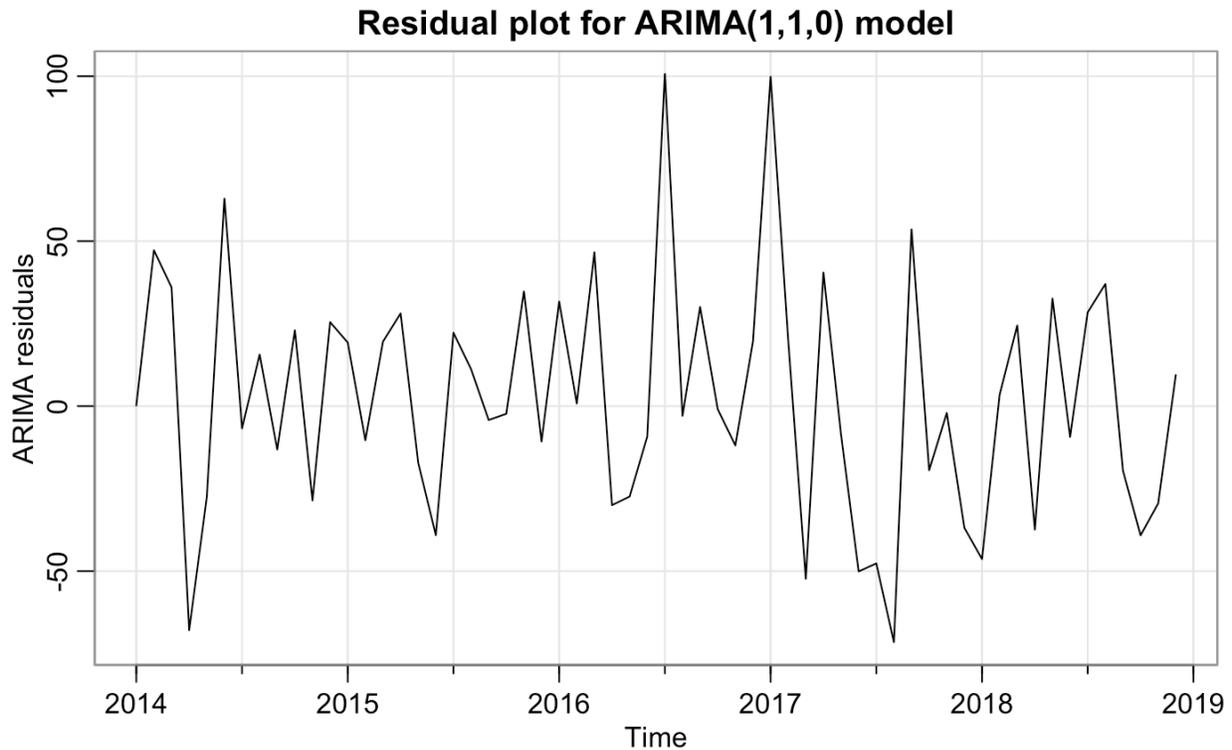

**Figure 8: Residual plot for Model 3.1, illustrating no further trend in the residuals.**

In conclusion, even when we account for the past via the inclusion of an ARIMA model, both weight and number of fentanyl seizures still have a statistically significant impact on overdose deaths. If one works at the level of counties, rather than at the state-wide level, then one must expand on Model 3.1 to account for autocorrelation among data points coming from the same county (Rosenblum et. al, 2020). For this, generalized linear mixed models are required (Wu, 2010). An even more advanced analysis is possible, using the techniques of topological data analysis to account for geographical proximity between counties.

## 6. Comparison between types of law enforcement

Sections 4 and 5 demonstrated that weight is a statistically significant predictor for overdose deaths, in both regression models and ARIMA models. In this section, we focus on the weights of drugs seized by different *types* of law enforcement. The goal is not to improve on

Model 3.1, but rather to understand which types of law enforcement are seizing which weights of drugs. Such information can be used when allocating funding to different types of law enforcement. We create four categories of law enforcement, and provide a table summarizing their activities:

(1) national level forces – including the FBI, DEA, ICE,

(2) specialized forces – including drug task forces and enforcement units, BCI, OOCIC,

(3) regular forces – including police departments and sheriff's offices, and

(4) unrelated forces that would not be included in our comparison (e.g. park districts, or drugs that were found abandoned).

|  | All drug seizures | Seizures involved fentanyl |
|---|---|---|
| National level forces | 1426 | 197 |
| Specialized forces | 18496 | 2082 |
| Regular forces | 99075 | 12372 |
| Unrelated forces | 2375 | 181 |

**Table 2: Summary statistics of number of drug seizures by categories of forces from 2014-2018**

The weight of the confiscated drugs is relevant for predicting how many lives are saved by these drug seizures. We begin with histograms comparing the first three categories of law enforcement, and immediately see that different types of law enforcement have strikingly different profiles of drug weights that they seize.

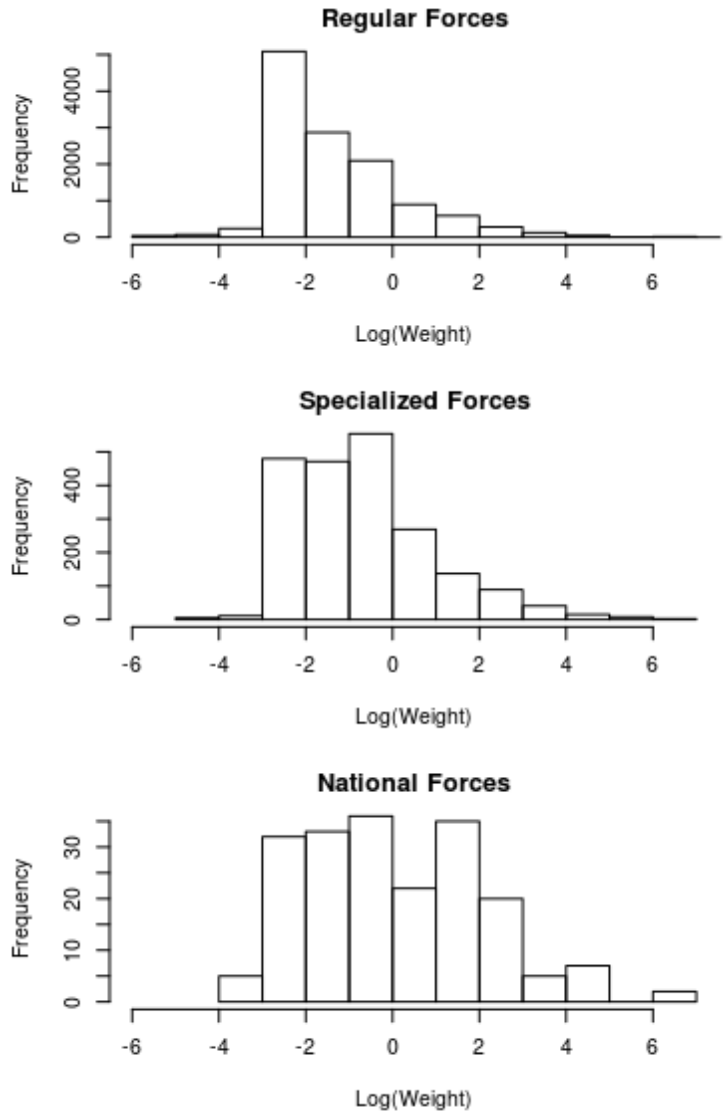

**Figure 9: The distribution of the log(Weight) of seizures by each category of force.**

In these histograms, we used the log of the weight of the confiscated drugs, because the distribution of the weight is extremely right-skewed with mostly small weight seizures and few extremely high weight seizures (Figure 10). This is due to the presence of some very large drug seizures of cannabis.

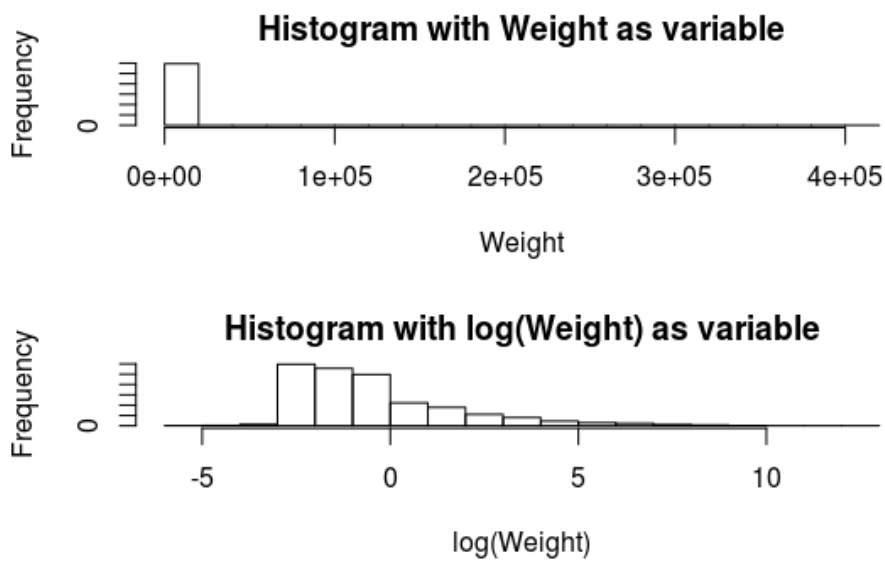

**Figure 10: Example between using Weight and log(Weight) as values in x-axis.**

The histograms in Figure 9 appear to represent different weight distributions: the regular forces had mostly small weight fentanyl-related seizures, while the specialized forces were better at seizing higher weights. On the other hand, the national level forces tended to get more medium and even high weight seizures compared to the other two forces, but many fewer seizures in general (Table 2). We can formalize the comparison of these distributions using the Kolmogorov-Smirnov test. This test compares two distributions: the null hypothesis states that the two distributions are the same, and the alternative hypothesis states that they are different. The test-statistic is the largest difference between the two cumulative density functions. This test statistic is distributed according to a chi-squared distribution (Massey, 1951). In our case, each of the three pairwise comparisons of the weight distributions yielded statistically significant differences (Kolmogorov-Smirnov: $p < 0.001$), confirming that different types of law enforcement really do have statistically significantly different seizure profiles. Even though low weight seizures are more predictive of deaths than high weight seizures, we hesitate to argue that regular forces are more effective than national forces at preventing drug overdose deaths, because there is value in removing fentanyl from the drug market at any stage of the pipeline. However, we do recommend low weight seizures be given higher priority for analysis by crime labs, that information about fentanyl prevalence be shared widely, and that law enforcement resources be directed towards life-saving harm reduction techniques whenever crime lab data reveals a larger than normal prevalence of fentanyl in low weight seizures.

We conclude this section with a brief discussion of ways this analysis could be improved. Firstly, the BCI dataset does not contain information about the amount of each detected substance, if there is a mixed seizure. Hence, we do not know how much fentanyl is present in a seized drug packet, though this information would help to paint a clearer picture of the prevalence of fentanyl in the drug supply. Secondly, it is possible that the tested sample is a proportion of the actual drug seizures, which means the true weight might be much larger. We encourage the BCI to report data to resolve these difficulties, so that a finer analysis may be conducted in a follow-up paper. We also call upon other crime labs to make their data available, as BCI has done, so that a better picture of the drug supply can be obtained.

**Summary of main conclusions**

In this section, we summarize the main conclusions of each of the sections above. Throughout, we have focused on the relationship between drug seizures containing fentanyl, and deaths due to opioid overdose. Our first model (Section 3) found that drug seizures in a given month explain 59% of the variability in deaths in the same month. Our next model (Section 4) improved on this by including variables keeping track of the number of seizures in different weight bins. We demonstrated that smaller weight seizures have a stronger effect on overdose deaths than larger weight seizures, and we obtained a model with an adjusted R-squared of

77.99%. Our next model (Section 5) extended on the previous two by factoring in the past, using time-series techniques. Specifically, we show that the number of overdose deaths in a given month depends on the number in the previous month, but even holding this relationship constant, both the number and the weight of drug seizures have a statistically significant effect on overdose deaths. Furthermore, we found that deaths are most highly correlated to immediate seizures (i.e., in the same month) rather than seizures in previous months. Hence, data on drug seizures can effectively alert law enforcement, harm reduction professionals, and drug users to the presence of fentanyl (or its analogues) in the drug supply. When fentanyl prevalence is high, harm reduction techniques such as naloxone, fentanyl test strips, and direct communication with drug users (e.g., at needle exchanges) should be deployed. As low weight seizures are most predictive of deaths, law enforcement should not ignore low weight seizures in favor of high weight seizures.

Lastly, in Section 6, we categorized the different types of law enforcement and demonstrated the weights of fentanyl-containing drugs that different types have been most effective in confiscating. We used a Kolmogorov-Smirnov test to demonstrate that the weight distributions seized by different types of law enforcement are statistically significantly different, as histograms also demonstratively show. Recognizing the power of drug seizure data for predicting overdose deaths, we call on crime labs (other than BCI) to release their data so that subsequent papers may extend on this one. We call on law enforcement and harm reduction professionals to target the most dangerous weights (i.e., low weight fentanyl) at high risk times (i.e., when fentanyl prevalence is high), and to use a variety of approaches to communicate with drug users and save lives.

**Conflict of interest**

Nothing to declare.

**Acknowledgments**

The authors wish to thank Dennis Cauchon and Harm Reduction Ohio for providing the BCI Drug Seizure dataset, as well as extremely valuable guidance. The authors thank the Denison University William G. Bowen and Mary Ellen Bowen Research Scholars fund for financing this research in the summer of 2019, Mine Dogucu for helpful conversations about statistical models, and the Denison Data Driven Journalism group (especially Jack Shuler and Anthony Bonifonte) for helping to shape our approach to the data.